\documentclass[twocolumn,showpacs,preprintnumbers,amsmath,amssymb,superscriptaddress, bibnotes]{revtex4}

\usepackage{graphicx}
\usepackage{dcolumn}
\usepackage{bm}

\begin{document}
\draft
\preprint{HEP/123-qed}

\title{Itinerant Nature of U~5$f$ States in Uranium Mononitride UN Revealed by Angle Resolved Photoelectron Spectroscopy}

\author{Shin-ichi~Fujimori}
\affiliation{Condensed Matter Science Division, Japan Atomic Energy Agency, Sayo, Hyogo 679-5148, Japan}

\author{Takuo~Ohkochi}
\altaffiliation[Present address: ]{Japan Synchrotron Radiation Research Institute/SPring-8, Sayo, Hyogo 679-5198, Japan}
\affiliation{Condensed Matter Science Division, Japan Atomic Energy Agency, Sayo, Hyogo 679-5148, Japan}

\author{Tetsuo~Okane}
\affiliation{Condensed Matter Science Division, Japan Atomic Energy Agency, Sayo, Hyogo 679-5148, Japan}

\author{Yuji Saitoh}
\affiliation{Condensed Matter Science Division, Japan Atomic Energy Agency, Sayo, Hyogo 679-5148, Japan}

\author{Atsushi~Fujimori}
\affiliation{Condensed Matter Science Division, Japan Atomic Energy Agency, Sayo, Hyogo 679-5148, Japan}
\affiliation{Department of Physics, University of Tokyo, Hongo, Tokyo 113-0033, Japan}

\author{Hiroshi~Yamagami}
\affiliation{Condensed Matter Science Division, Japan Atomic Energy Agency, Sayo, Hyogo 679-5148, Japan}
\affiliation{Department of Physics, Faculty of Science, Kyoto Sangyo University, Kyoto 603-8555, Japan}

\author{Yoshinori~Haga}
\affiliation{Advanced Science Research Center, Japan Atomic Energy Agency, Tokai, Ibaraki 319-1195, Japan}

\author{Etsuji~Yamamoto}
\affiliation{Advanced Science Research Center, Japan Atomic Energy Agency, Tokai, Ibaraki 319-1195, Japan}

\author{Yoshichika~\=Onuki}
\affiliation{Advanced Science Research Center, Japan Atomic Energy Agency, Tokai, Ibaraki 319-1195, Japan}
\affiliation{Graduate School of Science, Osaka University, Toyonaka, Osaka 560-0043, Japan}

\date{\today}

\begin{abstract}
The electronic structure of the antiferromagnet uranium nitride (UN) has been studied by angle resolved photoelectron spectroscopy using soft X-rays ($h\nu$=420-520~eV).
Strongly dispersive bands with large contributions from the U~5$f$ states were observed in ARPES spectra, and form Fermi surfaces.
The band structure as well as the Fermi surfaces in the paramagnetic phase are well explained by the band-structure calculation treating all the U~5$f$ electrons as being itinerant, suggesting that itinerant description of the U~5$f$ states is appropriate for this compound.
On the other hand, changes in the spectral function due to the antiferromagnetic transition were very small.
The shapes of the Fermi surfaces in a paramagnetic phase are highly three-dimensional, and the nesting of Fermi surfaces is unlikely as the origin of the magnetic ordering.

\end{abstract}

\maketitle
\narrowtext
\section{INTRODUCTION}
The origin of magnetism has been one of the controversial issues in $f$-electron materials.
Generally, in rare-earth 4$f$ compounds, a long range magnetic ordering is understood by the Ruderman-Kittel-Kasuya-Yoshida (RKKY) interaction, which is essentially based on a localized $f$ electron picture.
On the other hand, the origin of magnetism in actinide 5$f$ has been not well understood since the 5$f$ electrons show magnetic properties of both itinerant and localized.
Although there are a number of studies on the magnetism of actinide-based compounds, there are only few cases where the origin of magnetism is directly revealed\cite{UTe, UIr}.

In the present study, we report an angle resolved photoelectron spectroscopy (ARPES) study on uranium mononitride (UN) to understand its electronic structure as well as to explore the origin of magnetism in 5$f$ compounds.
UN has a NaCl type face center cubic (fcc) crystal structure in the paramagnetic (PM) phase, and it undergoes a transition into the type I antiferromagnetic (AFM) phase with $T_{\rm N}$=53~K and $\mu_{\rm ord}$=0.75~$\mu_{\rm B}$.
Although a band antiferromagnetism has been suggested for UN\cite{Fournier}, there are still controversial issues on the magnetic and electronic properties.

Samsel-Czeka\'{l}a {\it et al.} have studied the magnetic and transport properties of UN\cite{UN}.
They have discussed their data based on the dual and spin density wave (SDW) picture of the U~5$f$ states, but definitive conclusion was not obtained.
Solontsov and Silin \cite{weakAF} suggested that UN is a weak itinerant-electron antiferromagnet which has a different mechanism from the nesting of the Fermi surface.
They have suggested that the magnetic ordering is caused by the polarization of bands rather than the SDW type Fermi surface instability.
Therefore, it is essential to reveal the overall electronic structure of UN to understand the nature of the U~5$f$ states in UN.

The electronic properties of UN have been studied experimentally\cite{UN,Reihl,Itoh,UN_film,Lander} and theoretically\cite{Yin,UN_GGA}.
Reihl {\it et al.} first measured ARPES spectra of UN by using incident photon energy of $h \nu=$25~eV with the energy resolution of $\Delta E=$150~meV\cite{Reihl}.
They have observed the temperature dependence of the spectra, and suggested the itinerant nature of the U~5$f$ states.
However, they measured the normal emission spectrum only, and overall electronic structure of the U~5$f$ states were not well understood.
Afterward, Itoh {\it et al.} performed higher energy resolution ARPES experiments on UN by using $h\nu$=21.2~eV with $\Delta E=$50~meV, and observed two-dimensional energy band dispersions.
They observed two non-dispersive U~5$f$ bands in addition to dispersive N~$s$, $p$ bands\cite{Itoh}.
One is located just below $E_{\rm F}$, and the other is located at $E_{\rm B}$=0.6~eV.
Therefore, they have suggested that the U~5$f$ electrons have dual (itinerant and localized) natures.
Meanwhile, a recent XPS study on UN showed that the U~4$f$ core-level spectrum shows an asymmetric line shape characteristic of a metal and multiple final state structures, which were also interpreted as the dual nature of the U~5$f$ states.
From the theoretical point of view, it has been suggested that the electronic structure cannot be understood within the framework of a local density approximation (LDA).
Modak and Verma\cite{UN_GGA} studied the electronic structure of UN by the LDA as well as the generalized-gradient approximation (GGA), and suggested that the LDA is insufficient for the description of its electronic structure.
Moreover, Yin {\it et al.} \cite{Yin} theoretically calculated the thermal conductivity of actinide nitrides  by means of the dynamical mean field theory (DMFT), and pointed out that the electron correlation effect (the Hubbard $U$) is essential to describe its electrical properties.
Therefore, an appropriate theoretical framework is still not known for its description.

In addition to those scientific interests, there is a practical demand to study its electronic structure.
UN is a promising fuel material for the generation-IV advanced nuclear reactors since it has high melting point ($2850\ {}^\circ\mathrm{C}$), a very good thermal conductivity at high temperatures as well as a high fuel density (14.32 gcm$^{-2}$) \cite{UN_fuel}.
It is quite important to clarify its electronic structure to design better fuel materials\cite{Yin}.
Moreover, its understanding is essential to comprehend the reaction of UN with water or oxygen for the safety of nuclear power plants as well as for the storage of fuel materials\cite{UN_surface}.

\section{EXPERIMENTAL}
Photoemission experiments were performed at the soft X-ray beamline BL23SU of SPring-8 \cite{BL23SU}.
The overall energy resolution in the angle-integrated photoemission (AIPES) experiments at $h\nu $=800~eV was about 110~meV, and that in the ARPES experiments at $h\nu = 420-520$~eV was 80-120~meV depending on the experimental setup.
The position of the Fermi level ($E_{\rm F}$) was carefully determined by measurements of the evaporated gold film.
Clean sample surfaces were obtained by {\it in situ} cleaving the sample with the surface parallel to the (001) plane.
The position of ARPES cuts were calculated by assuming free-electron final states with the inner potential of $V_{\rm 0}$=12~eV.

\section{RESULTS AND DISCUSSION}
\subsection{Angle-integrated photoemission spectra}
\begin{figure}
\includegraphics[scale=0.4]{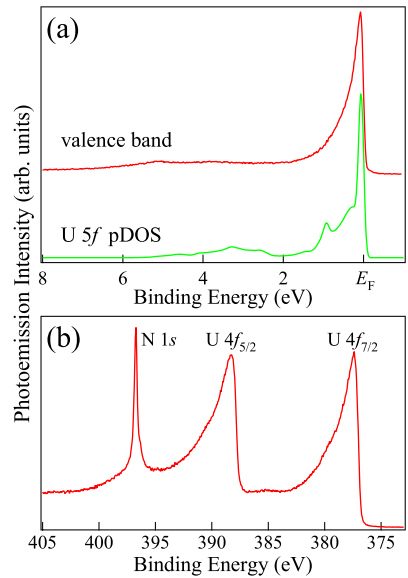}
\caption{Angle-integrated photoemission spectra of UN measured with $h\nu $=800~eV.
(a)Valence-band spectrum with the calculated U~5$f$ partial DOS broadened by the instrumental resolution 
(b)U~4$f$ core-level spectrum.}
\label{AIPES}
\end{figure}
First, we present the angle-integrated photoemission (AIPES) spectra of UN.
Figure~\ref{AIPES} (a) shows the valence band spectrum of UN taken at $h\nu $=800~eV.
The sample temperature was 75~K, and the compound is in the PM phase.
In this photon energy range, the contribution from the U~5$f$ states are dominant, and those from the N~$s$, $p$ states are two or three orders of magnitude smaller than that of the U~5$f$ states\cite{Atomic}.
In the valence-band spectrum, there is a sharp peak structure just below $E_{\rm F}$.
This peak structure has a strong contribution from the U~5$f$ states.
On the other hand, weak and broad peak structures distributed at 2-6~eV are ascribed to contributions from the N~$s, p$ states.
To compare with the experimental data, we have performed band-structure calculation treating all the U~5$f$ electrons as being itinerant.
The calculation is a relativistic-linear-augmented-plane-wave (RLAPW) band-structure calculations\cite{Yamagami} within a LDA\cite{LDA}
In this figure, the calculated U~5$f$ partial DOS (U~5$f$ pDOS) broadened with the instrumental resolution are also indicated.
It has an asymmetric line shape having a long tail towards higher binding energies.
The overall spectral line shape is consistent with the calculation.

Figure~\ref{AIPES} (b) shows the U~4$f$ core-level spectrum of UN taken at $h\nu $=800~eV.
It shows a spin-orbit splitting corresponding to U~4$f_{7/2}$ and U~4$f_{5/2}$, and both of them have a broad asymmetric line shape.
This is a common feature of U~4$f$ core-level spectra of metallic uranium compounds.
The spectrum shows a relatively simple main line shape with large asymmetry\cite{Ucore}.
This spectral line shape is similar to that of itinerant U~5$f$ compounds.
The binding energy of U~4$f_{7/2}$ main line is 377.27 eV, which is in a good agreement with the previous study on UN thin film (377.3 eV)\cite{UN_film, PuN_film}.
This is within the binding energies of 5$f^4$ final-state peak of various itinerant uranium compounds\cite{Ucore}, suggesting that UN can be classified into itinerant uranium compounds.
In the previous photoemission experiment, a small satellite structure was observed at 3~eV higher binding energy side than the main lines\cite{UN}.
They have argued that this satellite originates from different valence states, and suggested this is an indication of the dual nature of the U~5$f$ states in this compound.
However, there is no such satellite structure in the present U~4$f$ spectrum.
Since the peak position of the small satellite corresponds to that of UO$_2$, this satellite may originate from oxidized component of their sample surfaces.

\subsection{Band structures}
\begin{figure*}
\includegraphics[scale=0.2]{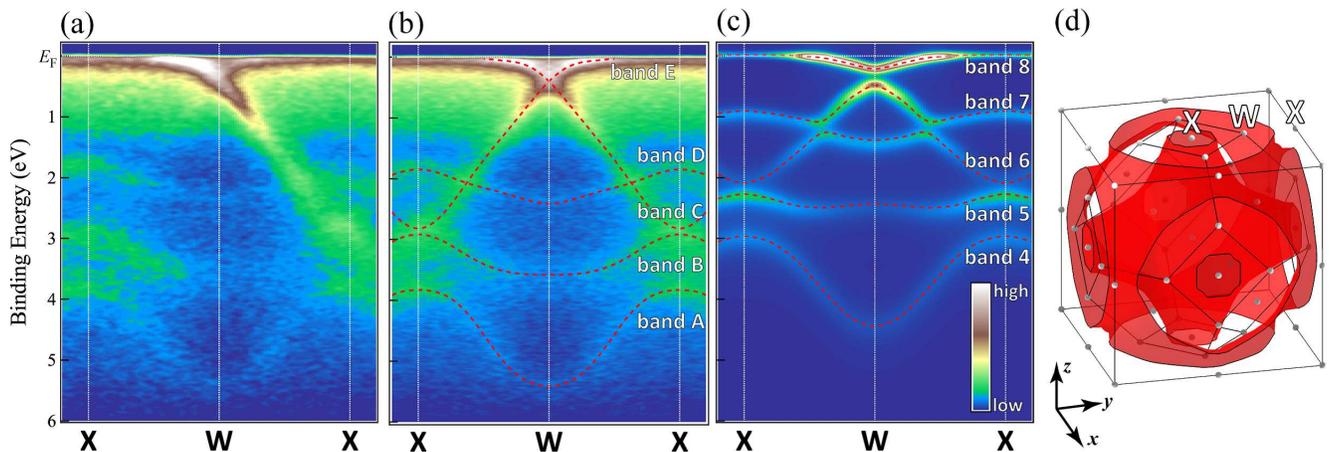}
\caption{(Online color)ARPES spectra and their comparison with the result of band-structure calculation.
(a)ARPES spectra measured along the X-W-X line.
(b)Symmetrized ARPES spectra. The dashed curves are guide to the eye.
(c)Simulation of ARPES spectra based on the band structure-calculation treating all the U~5$f$ electrons as being itinerant.
(d)fcc Brillouin zone of UN in the paramagnetic phase and calculated Fermi surfaces.}
\label{ARPES_XWX}
\end{figure*}
Figure~\ref{ARPES_XWX} (a) shows the ARPES spectra of UN measured along the X-W-X line.
The sample temperature was kept at 75~K in the PM phase.
The position of the ARPES cut in momentum space is calculated based on the free electron final sates, and the photon energy used was 490~eV.
The experimental energy resolution was 85~meV.
In the ARPES spectra, clear energy dispersions were observed.
In the vicinity of $E_{\rm F}$, there exists a strongly dispersive band with strong intensity.
This band has a large contribution from the U~5$f$ states especially in the vicinity of $E_{\rm F}$.
As the band goes from the W point to the X point in the first Brillouin zone, it approaches $E_{\rm F}$.
Near the midpoint of the X-W line, its intensity suddenly decreases, suggesting that it crosses $E_{\rm F}$.
Meanwhile, weak but finite photoemission intensities just below $E_{\rm F}$ persist outside the Fermi momentum ($k_{\rm F}$).
A similar phenomenon has been observed in the ARPES spectra of other uranium compounds\cite{UB2_ARPES}.
Its origin shall be discussed below.
On the high binding energy side ($E_{\rm B}=1.5-6$~eV), there exist weak and strongly dispersive bands.
Since they have weaker intensities than those of bands near $E_{\rm F}$, they are assigned to contributions mainly from the N~$s, p$ bands.
The overall structure of the present spectra are very similar to the previous results measured by He I ($h\nu=$21.2~eV)\cite{Itoh}, but there is one striking difference between them.
In the previous ARPES study, two non-dispersive bands were observed in the vicinity of $E_{\rm F}$\cite{Itoh}.
One is located just below $E_{\rm F}$, while the other is located around $E_{\rm B} \sim$0.6~eV, and they were assigned to the itinerant and localized components, respectively.
This has been considered as the indication of the dual nature of the U~5$f$ electrons in UN.
However, we have observed a single itinerant band in the vicinity of $E_{\rm F}$, and the dual nature of the U~5$f$ electrons was not observed.
Those differences may originate from the higher surface sensitivity in the previous ARPES study.

Here, it should be noted that the intensities of observed bands are not symmetric with respect to the W point.
For example, the band located just below $E_{\rm F}$ has an asymmetric shape relative to the W point, and the intensity of its counterpart is very weak.
Since their intensities depend on the Brillouin zone, this may be due to the photoemission structure factor (PSF) effect as has been observed in ARPES spectra of other materials \cite{PSF}.
To eliminate this effect, we have symmetrized ARPES spectra relative to the W point as shown in Fig.~\ref{ARPES_XWX} (b).
Dashed curves represent approximate positions of bands, estimated from the second derivatives of ARPES spectra.
The band structure of UN is more easily understood from this image.
There are five bands in this energy-momentum region, and they are named as A, B, C, D, and E from higher to lower binding energies.
The Band E forms an electron pocket around the W point, and they have a large contribution from the U~5$f$ states.
Bands A-D on the high binding energy side disperse strongly, and are assigned to contributions mainly from the N~$s$ and $p$ states.

To understand the validity of the itinerant description of the 5$f$ states in this compound, we compare the present ARPES spectra with the result of the band-structure calculation within the LDA framework treating all the U~5$f$ electrons as being itinerant.
Figure~\ref{ARPES_XWX} (c) shows the band-structure and a simulation of ARPES spectra based on the band-structure calculation, and Fig.~\ref{ARPES_XWX} (d) shows the fcc Brillouin zone of UN with the calculated Fermi surfaces.
In this simulation, the following effects have been taken into account; (i) the broadening in the $k_z$ direction due to the finite escape depth of photoelectrons, (ii) the lifetime broadening of photo-hole, (iii) photoemission cross-sections of orbitals, and (iv) the energy resolution and angular resolution of the electron analyzer.
Their details are described in Appendix A.
Energy band dispersions corresponding to the X-W-X high symmetry line are shown in dashed curves, and are consistent with the results of the previous band structure-calculations\cite{UN_calc1, UN_calc2}.
In the band-structure calculation, Band 8 forms the hole pocket Fermi surface around the W point.
On the high binding energy side, there exist strongly dispersive bands with contributions mainly from the N $s$ and $p$ states.
A comparison between ARPES spectra and the simulation shows that there is a good correspondence between the experimentally observed bands A-E and the calculated bands 4-8, respectively, though the binding energies of bands A-D are deeper in the experiment than in the calculation.
On the other hand, there are some disagreements between the experiment and the calculation.
For example, there is a clear gap structure between the bottom of band 8 and the top of band 7 in the band-structure calculation, while they are not clearly seen in experiment.
Bands 7 and 8 are mainly consist of the N~$s, p$ states and U~5$f$ state, respectively, and the absence of a clear gap in the experimental spectra suggests that the hybridization between them is weaker in the experiment than in the calculation.
Although there exist those discrepancies, a good one-to-one correspondence between them suggests that band-structure calculation gives a reasonable description of the experimentally obtained band structure.
Here, it should be noted that there exist finite photoemission intensities at $E_{\rm F}$ outside $k_{\rm F}$ in this simulation as it was observed in the experiment.
Its origin is mainly due to the finite broadening along momentum directions and the three dimensionality of its electronic structure.
Therefore, these finite intensities at the Fermi energy observed in the ARPES spectra of other uranium compounds such as UB$_2$ would be explained by this effect.

\subsection{Temperature dependence of band structure}
\begin{figure*}
\includegraphics[scale=0.15]{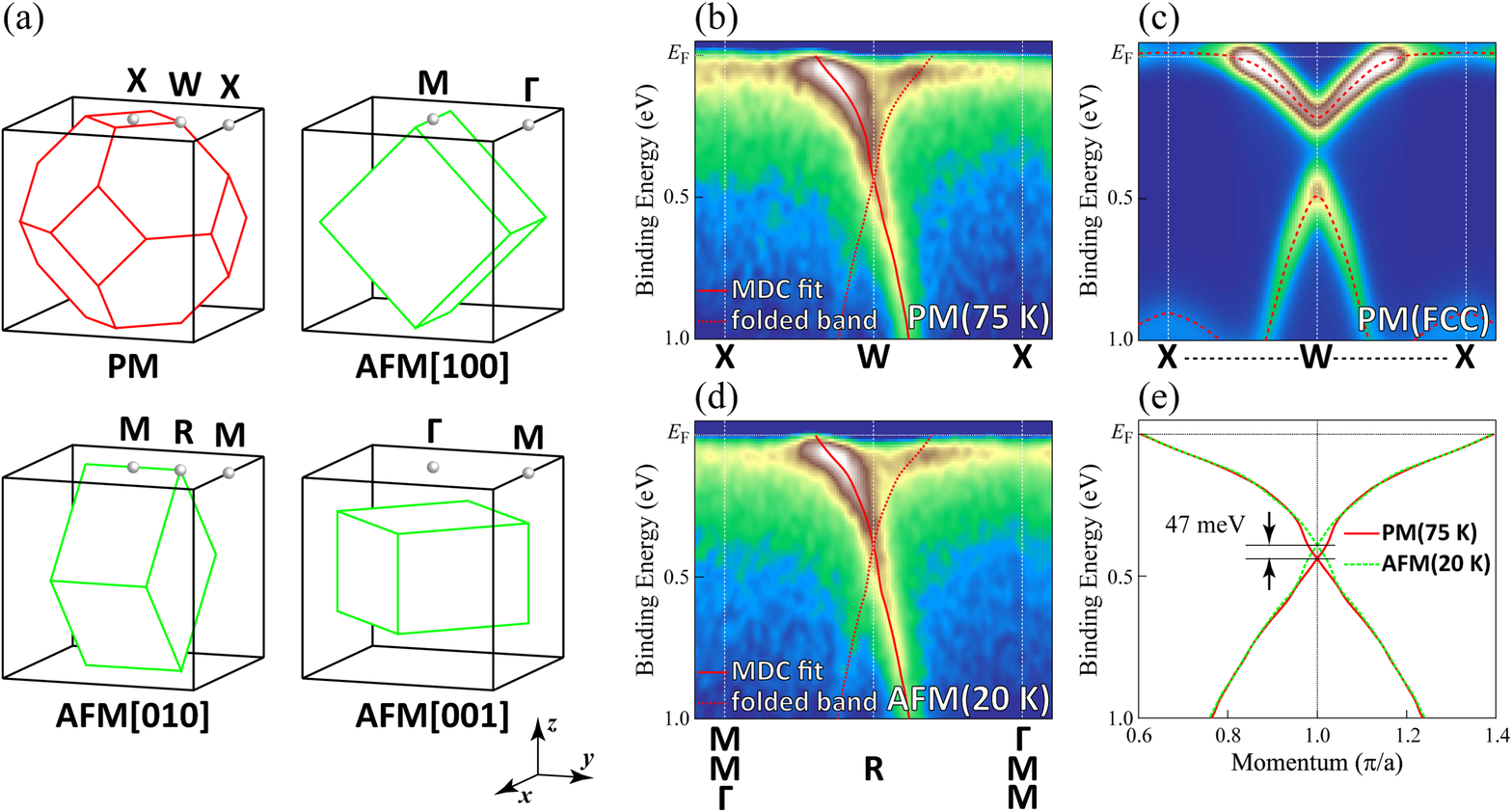}
\caption{(Online color)Temperature dependence of ARPES spectra.
(a) Brillouin zone of UN in the PM phase and in the AFM phase with the AFM ordering wave vector along [100], [010], and [001] directions.
(b) ARPES spectra of UN in the PM phase measured along the X-W-X direction.
(c) Simulation of ARPES spectra based on band-structure calculation of UN in the PM phase treating all the U~5$f$ electrons as being itinerant.
(d) ARPES spectra of UN in the AFM phase measured along the same direction as the scan in the PM phase.
The scan corresponds to M-$\Gamma$ , M-R-M, and $\Gamma$-M directions in the AFM Brillouin zone.
(e) Comparison of bands near $E_{\rm F}$ in the PM and AFM phases.
}
\label{ARPES_tmp}
\end{figure*}
Next, we show the changes of the electronic structure associated with the antiferromagnetic transition.
Before we show the ARPES spectra, we explain a relationship between the ARPES scan in the PM phase and the AFM phase.
In the PM phase, the crystal structure is fcc.
In the AFM phase, the magnetic moments are aligned ferromagnetically within the (001) plane, and are coupled antiferromagnetically between the neighboring (001) plane.
There are three equivalent directions of the AFM ordering vector, [100], [010], and [001], and the Brillouin zone should be a diamond shape whose direction depends on the directions shown in Fig.~\ref{ARPES_tmp} (a).
Since those ordering directions will form domains on a very small length, and the beam spot is expected to cover multiple domains.
The ARPES spectra taken along the X-W-X direction in the PM Brillouin zone corresponds to a mixture of signals from the M-$\Gamma$, M-R-M, and $\Gamma$-M directions in the AFM phase.
The relationship between the PM and the AFM Brillouin zone is as shown in Fig.~\ref{ARPES_tmp} (a).

Figure~\ref{ARPES_tmp} (b) and (c) show a blowup of the near $E_{\rm F}$ region of ARPES spectra measured along the X-W-X direction and the calculated energy band dispersions in the PM phase.
The positions of bands estimated from their momentum distribution curves (MDCs) and its folding are also shown in Fig.~\ref{ARPES_tmp} (b) by solid and dashed curves, respectively.
The behavior of the quasi-particle band is more clearly recognized.
Around the W point, there is a V-shaped band just below $E_{\rm F}$, and it forms an electron pocket Fermi surface.
Here, it should be noted that the fitted band has a larger slope than the calculated band in the vicinity of $E_{\rm F}$.
This seems to be inconsistent with the fact that the experimental band has a heavier electron mass than that of the band-structure calculation as it is inferred from the larger electronic specific heat coefficient in the experiment ($\gamma_{\rm e} =$49~mJK$^{-2}$mol$^{-1}$) than in the calculation ($\gamma_{\rm e} =$17.95~mJK$^{-2}$mol$^{-1}$).
This might be mainly due to the very small energy scale of the renormalization of the experimental quasi-particle bands.
In fact, the effect of the renormalization appears in the energy range of few tens meV of the vicinity of $E_{\rm F}$ in the quasi-particle bands of heavy fermion compound USb$_2$\cite{USb2_kink}.
If the quasi-particle band has a renormalization in a similar energy scale, the structure cannot be observed with the present experimental energy resolution.
Moreover, a fitting of MDCs generally gives a larger slope than the actual band when the band is nearly flat and broad\cite{EDCMDC}.
This can be inferred from Figs~\ref{ARPES_tmp} (b) and (d) where fitting of energy distribution curves (EDCs) should give nearly flat band just below the $E_{\rm F}$ around $k_{\rm F}$.
Therefore, the peak positions estimated from MDCs in the vicinity of $E_{\rm F}$ do not correspond to the actual band positions in the present analysis.

Figure~\ref{ARPES_tmp} (d) shows the ARPES spectra measured by the same geometry of Fig.~\ref{ARPES_tmp} (b), but at 20~K.
The sample is in the AFM phase, and this scan direction would correspond to the M-$\Gamma$, M-R-M, and $\Gamma$-M directions in the AFM Brillouin zone.
It is shown that spectra do not show significant changes.
Here we note that the N~$s,p$ derived bands located in binding energy range $E_{\rm B} = 1.5 - 6$~eV, show no changes by the AFM transition.
Back-folded replica bands due to the AFM Brillouin zone are not clearly observed.
The MDC-fitted peak positions and its folding are shown by solid and dashed curves.
The basic structure is essentially identical to those measured in the PM phase.
Therefore, the changes in ARPES spectra associated with AFM transition is very small.

To see the changes of those bands in detail, we have compared fitted bands in both phases.
Figure~\ref{ARPES_tmp} (e) shows a comparison of the fitted bands between the PM and AFM phases.
Since those bands should be symmetric with respect to $k_x = \pi/a$, folded bands are also shown in this figure.
The bands are almost identical in the PM and the AFM phases, but there exists a small but clear difference around the zone boundary ($k_x \sim \pi / a$).
As the compound undergoes the AFM phase transition, the crossing point of bands moves toward lower binding energies by about 47~meV.
Since the electronic structure of this crossing point is affected by the folding of bands due to the magnetic ordering [010], this change is considered to be due to the PM to AFM transition.
We shall consider the origin of this change in the discussion below.

\subsection{Fermi surface}
\begin{figure}
\includegraphics[scale=0.2]{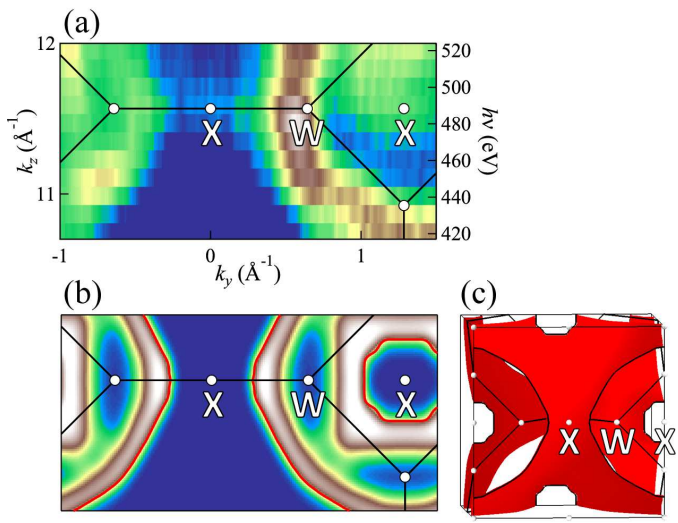}
\caption{(Online color)Fermi surfaces of UN.
(a)Experimental Fermi surface mapping in the $k_y$-$k_z$ plane obtained by changing the incident photon energy.
(b)Simulation of Fermi surface mapping based on band-structure calculation treating all the U~5$f$ electrons as being itinerant.
(c)Brillouin zone in the PM phase and calculated Fermi surfaces.}
\label{ARPES_FS}
\end{figure}
To reveal the overall shapes of the Fermi surfaces of UN in three-dimension, we have performed Fermi surface mapping by changing photon energies.
Figure~\ref{ARPES_FS} (a) shows an intensity map of ARPES spectra obtained by changing the photon energy from 420 eV to 520 eV.
The sample temperature was kept at 20~K, and the sample was in the AFM phase.
Photoemission intensities within $E_{\rm F} \pm 50$~meV of each ARPES spectra were integrated, and mapped as a function of momenta parallel ($k_y$) and perpendicular ($k_z$) to the sample surface.
A round-shaped FS around X point is observed.
Figure~\ref{ARPES_FS} (b) and (c) show the simulation of FS mapping and three dimensional shape of FS calculated by the band-structure calculation, respectively.
Although the sample is in the AFM phase, the essential band structure near $E_{\rm F}$ does not show significant changes as shown in Fig.~\ref{ARPES_tmp}, we have compared the experimental Fermi surfaces with the band-structure calculation in the PM phase.
The large and round-shaped FS centered at the X point was observed while a small and square shaped FS centered at the X point was not clearly observed although both of them originate from the same band.
Those are due to the PSF effect as it was observed in the ARPES spectra shown in Fig.~\ref{ARPES_XWX} (a).
The cross-sections of the Fermi surfaces are also shown by the red solid curves.
Here, it should be noted that the photoemission intensities become strong outside and inside the round-shaped and square-shaped Fermi surfaces, respectively, in the simulation.
This is due to the finite energy resolution ($\sim$85~meV) as well as finite energy window ($\sim$100~meV) of photoemission intensity integration, which make the image include intensities from bands below $E_{\rm F}$.
Although the size of experimentally observed Fermi surface is slightly smaller than that of the calculation, the shape of large round Fermi surface centered at the X point matches between experiment and calculation.
Therefore, the experimental Fermi surfaces are well explained by the band-structure calculation.
This result again suggests that itinerant description is appropriate for the electronic structure of UN.

\subsection{Discussion}
As described above, we have observed the itinerant nature of the U~5$f$ electrons in UN.
The dual nature of the 5$f$ electrons was not observed in the present experiment, and the itinerant description is the most realistic starting point to describe the electronic structure of UN.
This suggests that the magnetic ordering in UN originates from the itinerant U~5$f$ electrons.
The magnitude of magnetic moment in the band structure calculation in the AFM phase is 0.50 $\mu_{\rm B}$ while the experimental value is 0.75 $\mu_{\rm B}$.
They are similar magnitude, and this also supports an itinerant description of U~5$f$ states in UN.
Here, it should be noted that the magnetic susceptibility in the PM phase follows a modified Curie-Weiss (CW) law above $T_{\rm N}$ with the effective moment of 2.65~$\mu_{\rm B}$\cite{UN}.
Therefore, the CW behavior in the PM phase also originates from itinerant U~5$f$ electrons.
Meanwhile, the changes of ARPES spectra due to the AFM transition were very small.
In addition, the shape of the Fermi surface of UN has highly three-dimensional shape, and its nesting region is very small, suggesting that the nesting of Fermi surfaces is unlikely as an origin of the AFM transition.
Therefore, although the itinerant description is appropriate for the U~5$f$ states, a simple nesting scenario cannot be applied to the AFM ordering in UN.
This aspect is consistent with the picture of the weak itinerant antiferromagnetism\cite{weakAF} where the formation of gap at $E_{\rm F}$ is not the main origin of the magnetism.

Here, we consider the changes of ARPES spectra associated with the AFM transition.
The changes of spectral function due to the AFM transition have been studied for some itinerant antiferromagnet experimentally\cite{Cr_ARPES1,Cr_ARPES2,Cr_ARPES3,FeAs_ARPES}.
The most extensively studied material is chromium metal, which is an itinerant antiferromagnet showing incommensurate SDW-type ordering.
In the ARPES studies of chromium metal, the AFM transition was observed in the ARPES spectra as the emergence of back-folded replica bands due to the magnetic Brillouin zone, and the formation of a hybridization gap in a large portion of the Fermi surfaces.
The back-folded bands are hybridized with the original bands at the boundary of the magnetic Brillouin zone, and the intensity is transferred from the original bands to the back-folded bands in the vicinity of their crossing points.
The spectral intensities of the upper and lower split bands are given by the coherent factors $u_p^2$ and $v_p^2$, respectively \cite{ARPES_AF1, ARPES_AF2}.
For UN, the formation of the hybridization gap nor the back-folded bands were not so clear as chromium metal at the zone boundary of the magnetic Brillouin zone.
Meanwhile, the position of the band is shifted toward lower binding energies by about 47~meV at the boundary of the magnetic Brillouin zone in which the magnetic moment is directed along the [010] direction as shown in Fig.~\ref{ARPES_tmp} (e).
This might be due to the formation of a small hybridization gap in the vicinity of their crossing point.
The small gap results from the small hybridization, and the folded bands should have a weak intensity in this case.
Therefore, the observation of the hybridization gap itself was impossible in the present experiment, but the observed changes might be due to the formation of the gap.
The small changes of electronic structure due to the AFM transition are consistent with the picture of the weak itinerant antiferromagnetism where spin-polarized itinerant electrons form moments in both of the PM and the AFM phases.

\section{CONCLUSION}
In conclusion, we have revealed the band structure and Fermi surfaces of UN by soft X-ray ARPES.
Bands with large contribution of the U~5$f$ states form clear bands in the vicinity of $E_{\rm F}$, and the dual nature of the U~5$f$ was not observed.
Both the band structure and the Fermi surfaces in the PM phase were qualitatively explained by the band-structure calculation based on the LDA which treats all the U~5$f$ electrons as being itinerant.
The dual nature of the U~5$f$ electrons was not observed in the present experiment, and the LDA is a realistic starting point to describe the electronic structure of UN.
Meanwhile, the changes of ARPES spectra associated with AFM transition were very small.
The Fermi surfaces of UN have highly three dimensional structures, suggesting that the nestings of Fermi surface are unlikely to be the origin of the AFM ordering.

\acknowledgments
We thank H.~Kusunose of Ehime University and K.~Shimada of Hiroshima University for helpful discussions.
The experiment was performed under the Proposal No. 2006B3808 at SPring-8 BL23SU.
The present work was financially supported by a Grant-in-Aid for Scientific Research from the Ministry of Education, Culture, Sports, Science, and Technology, Japan under contact No.21740271, a Grant-in-Aid for Scientific Research on Innovative Areas "Heavy Electrons" (No. 20102003) of The Ministry of Education, Culture, Sports, Science, and Technology, Japan, and the Shorei Kenkyuu Funds from Hyogo Science and Technology Association.

\appendix
\section{Simulation of ARPES spectra based on band-structure calculation}
\begin{figure}
\includegraphics[scale=0.08]{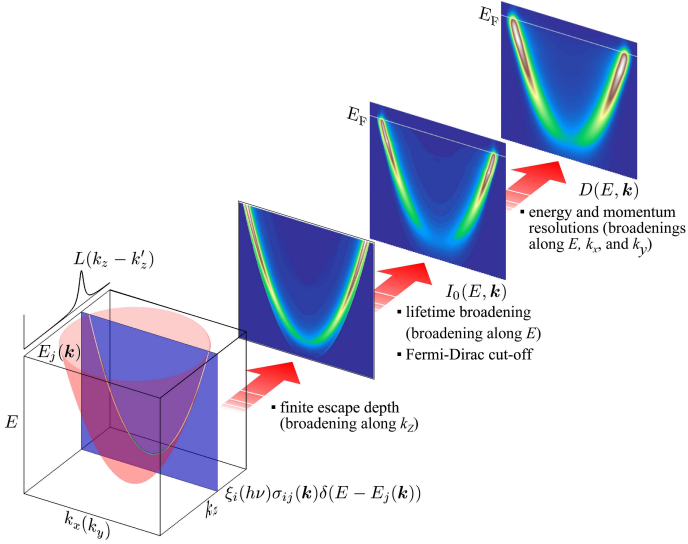}
\caption{(Online color)Simulation of ARPES spectra based on band-structure calculation.}
\label{ARPES_sim}
\end{figure}
In the present study, we have simulated ARPES spectra based on results of band-structure calculations.
Figure~\ref{ARPES_sim} shows the procedure of the simulation.
First, we take into account the intrinsic effects of photoemission process such as the damping of the final state wave functions and the finite life time of the photo-hole\cite{Wadati}.
Those effects appear as a finite broadening of photoemission spectra along the momentum perpendicular to the surface direction ($\delta k_z$) and energy direction ($\delta E$), respectively.
The matrix-element effect is approximately taken into account by multiplying the calculated photoionization cross section of atomic orbital $\xi_i(h \nu)$\cite{Atomic} and orbital character of each eign value in the band-structure calculation ${\sigma_{ij} (\mbox{\boldmath $k$)}}$ where $i$ and $j$ represent indices of orbital and band, respectively.
Based on those assumptions, the photoelectron current $I(E,\mbox{\boldmath $k$})$ is proportional to $I_0(E,\mbox{\boldmath $k$})$ expressed as
\begin{widetext}
\begin{equation}
I(E,\mbox{\boldmath $k$}) \propto
I_0(E,\mbox{\boldmath $k$}) = f(E) \sum_{i} \sum_{j}
\int_{-\infty}^\infty dk'_z  \xi_i(h \nu) \sigma_{ij} (\mbox{\boldmath $k$}) 
\frac{\delta k_z}{(k_z - k'_z)^2 + (\delta k_z / 2)^2}
 \frac{ \delta E}{(E - E_j(\mbox{\boldmath $k$}))^2 + (\delta E / 2)^2},
\end{equation}
\end{widetext}
where $E_j(\mbox{\boldmath $k$})$ is the energy dispersion of the $j$-th band.
This equation corresponds to Eq.~(3) of Ref.~\cite{Strocov} where we have further taken into account the contribution from Fermi-Dirac function and assumed that final-state surface transmission $|T^f|$ is constant.
Then, we take into account extrinsic effect such as an instrumental energy resolution ($\Delta E$) and momentum resolutions along parallel to the sample surface ($\Delta k_x$ and $\Delta k_y$).
The photoelectron current is proportional to $D(E,\mbox{\boldmath $k$})$ defined by
\begin{widetext}
\begin{equation}
D(E,\mbox{\boldmath $k$}) = 
\int_{-\infty}^\infty dE' G (E-E') \Bigl|_{\Delta = \Delta E}
\int_{-\infty}^\infty dk'_x G (k_x-k'_x) \Bigl|_{\Delta = \Delta k_x}
\int_{-\infty}^\infty dk'_y  G (k_y-k'_y) \Bigl|_{\Delta = \Delta k_y}
I_0(E,\mbox{\boldmath $k$}),
\end{equation}
\end{widetext}
The Gaussian broadening function $G_{\Delta} (x)$ is expressed as

\begin{equation}
G (x) = \frac{1}{\Delta} \sqrt{\frac{\ln 2}{\pi}}\exp \biggl\{ -{\frac{x^2}{(\Delta/2)^2}}\biggr\},
\end{equation}

where $\Delta / 2 \sqrt{ \ln{2} }$ is the FWHM width.
In the present study, the contributions from the U~5$f$, N~2$s$, and N~2$p$ states are taken into account since ionization cross sections of the other orbitals are much smaller than the values of those orbitals.
The ratios of photoionization cross sections are taken as $\xi_{{\rm U 5}f}$:$\xi_{{\rm N 2}s}$:$\xi_{{\rm N 2}p}$ = 1:0.2:0.04 based on the calculated cross sections of atomic orbitals\cite{Atomic}.
The broadening along the $k_z$ direction is assumed to be 0.1~\AA$^{-1}$, which corresponds to the escape depth of photoelectrons as $\lambda = 10$~\AA.
Life time broadening $\delta E$ is proportional to $(E-E_{\rm F})^2$ near $E_{\rm F}$ (an order of few tens to few hundred meV) for interacting Fermion system\cite{Luttinger} in the absolute zero temperature, but its behavior far below $E_{\rm F}$ (an order of several eV) is not well understood.
Thus, we have assumed that it has a linear dependence on $(E-E_{\rm F})$ on a wide energy scale as has been observed in Ni metal\cite{Ni}.
We have assumed that it is zero at $E_{\rm F}$ and 0.5 eV at $E_{\rm B}=5$~eV.


\end{document}